# Beams of particles and papers

## How digital preprint archives shape authorship and credit

Alessandro Delfanti

University of Toronto, Canada

**Abstract** In high energy physics, scholarly papers circulate primarily through online preprint archives based on a centralized repository, arXiv, that physicists simply refer to as 'the archive'. This is not just a tool for preservation and memory, but also a space of flows where written objects are detected and their authors made available for scrutiny. In this work I analyze the reading and publishing practices of two subsets of high energy physicists: theorists and experimentalists. In order to be recognized as legitimate and productive members of their community, they need to abide by the temporalities and authorial practices structured by the archive. Theorists live in a state of accelerated time that shapes their reading and publishing practices around precise cycles. Experimentalists turn to tactics that allow them to circumvent the slowed-down time and invisibility they experience as members of large collaborations. As digital platforms for the exchange of scholarly articles emerge in other fields, high energy physics could help shed light on general transformations of contemporary scholarly communication systems.

In high energy physics (HEP), the spheres in which information circulates and credit is attributed are partially separated from the ones formally used to evaluate research output. Scholarly papers circulate primarily in venues outside the traditional peer-reviewed scholarly journals. The community exchanges, retrieves, and evaluates papers almost exclusively through online preprint or 'e-print' archives that publish research manuscripts that have not yet been subject to a formal peer review and editorial process. These services are based on a centralized open access repository, arXiv.org, that physicists simply refer to as 'the archive', and other websites that use its data to provide more complex information about authors or citations and are collectively referred to as 'archives'. In sum, most HEP physicists do not read journals in their field but rather access written content through arXiv or other repositories that capture the articles deposited on it. Furthermore, epistolary and face-to-face communication are key in processes of information exchange and individual evaluation. However, HEP physicists do publish their work in journals and spend considerable money and effort in sustaining them. This separation between what is read and formal peer-reviewed publications makes HEP an ideal case for investigating the relation between the scholarly communication system and the dynamics of scientific credit and attribution. HEP's reading and publishing practices diverge even from those of other subfields within physics. Yet studying its archive could help understand more general transformations of contemporary scholarly communication systems. HEP physicists need to master the functioning of their archives in order to be claim credit for what they have done and signal their presence as productive and legitimate members of their communities. Negotiations about what constitutes a publication, who is an author, and which temporalities are to be met on arXiv point to problems that are increasingly common for other scholarly fields.

In the past few years, platforms for the publication of scholarly articles beyond or collaterally to peer-reviewed journals have proliferated, taking a variety of forms. They are now available in many sectors of research and attract millions of researchers. Some are public, some are private, others are relatively low-tech, whereas others have adopted technologies that are typical of the most recent social media. Created in 2013 and explicitly inspired by the physics repository, biorXiv.org provides a self-publishing platform for the life sciences. In fields in the social sciences and humanities, commercial services such as academia.edu have been able to supersede institutional and public archives and have attracted increasing volumes of preprints. In 2016, the for profit academic publisher Elsevier has entered the field by purchasing SSRN, a repository that is increasingly popular with law and economics scholars.

Disciplines in which researchers have been using open archives for decades as the main infrastructures to exchange their own papers – such as high energy physics – show that these instruments are connected to specific ways of circulating scientific writing and structuring the scientific community. Preprint archives are closely connected to the existence of what has been called a 'preprint culture', developed and settled since the early 1960s (Bohlin, 2004; Mele et al., 2006). Disciplines either not having developed this culture or having epistemic, institutional or social structures different from physics have been facing specific issues and may struggle to adopt preprint archives as the central place in which knowledge is shared and debated.

High energy physics' use of arXiv as the core communication platform is often used as an example of the evolution of open access in the era of digital scholarly communication. For example, Fitzpatrick points to this platform as having 'in effect replaced journal publication as the primary mode of scholarly communication' within some scientific fields (2011: 23; see also Borgman, 2007). Yet there seems to be a confusion about the role and significance of HEP preprint archives. Are they tools for access, preservation, discussion, curation or peer review? arXiv has been described as one of the pieces of evidence that point to a future death of the peer-reviewed paper, while at the same time underpinning knowledge circulation, review and preservation. Yet a closer look at the roles of preprints in HEP physics allows for a rethinking of such claims and urges a more nuanced analysis of these spaces. First, these repositories exist alongside a system of peer-reviewed journals that provide legitimacy to part of the content circulated on archives. While high energy physicists read and even cite arXiv papers, formal journal publications respond to the need for recognition and assessment by groups outside of the strict HEP community, such as funders or colleagues in collateral fields (Gunnarsdottir, 2005; Larivière et al., 2014). Second, archives are the spaces where the boundaries of the HEP field are continuously created and maintained. Publication on arXiv functions as a form of certification that separates insiders from outsiders (Reyes-Galindo, 2016). In this paper I focus on individual practices in order to add yet another layer to the picture. I contend that by mastering the sociotechnical characteristics of preprint archives, high energy physicists publicly affirm their existence as individuals within their community.

But are scholarly articles important at all for HEP physicists? In one of the most important works on high energy physics culture, Traweek (1988) denies the importance of publications. Indeed, Traweek states that what has been written and published is uninteresting to physicists: Published work is uncontested and does not get read. High energy physicists, according to Traweek, rarely read journals or visit the library but 'do scan preprints in order to know who is writing about what', (1988: 121). To achieve this goal they also rely on what Traweek calls 'gossip' (word of mouth communication), as physicists want to know what is going on before a piece of research is published. Oral communication conveys restricted information while written works contain public information, which is made accessible to the whole community but might be old, incomplete or unimportant. In Traweek's work, articles and preprints assume the role of commodities exchanged among physicists in order to claim ownership and 'pay royalties' to colleagues in form of citations. Written articles and preprints are thus record-keeping devices that are merely a trace of something happening elsewhere – Traweek (1988: 122) uses the metaphor of the traces left by particles in a bubble chamber, a technology based on liquid hydrogen invented in the 1950s and used in modern physics to detect electrically charged particles moving through it.

With this paper I aim at adding a piece to the puzzle, showing that while 'gossip' might still be one of the main communication practices of the physics community, the digitization of the archives has brought about a change in physicists' relation to written communication. Today, archives are crucial spaces that are central to the HEP community. Beware: I do not maintain that the *content* of HEP physicists' papers has acquired importance because of the transposition of archives onto digital technologies. Communication on 'physics', as physicists would describe the kind of scientific knowledge they see as appropriate for discussion amongst members of their community, might still happen in face-to-face communication, for example in conferences or at the chalkboard, and epistolary or personal interactions. Today, this is often mediated by the digital communication technologies and thus takes the form of webinars, Skype calls or emails. Nevertheless, the growth of HEP into a transnational enterprise and the digitization of its communication system have modified this picture. Preprint archives have thus assumed a central position that goes beyond their role as record-keeping devices or certification sites.

**The archive as a detector**

Here I stress the temporal and spatial dimensions of arXiv. In each category, corresponding to sub-fields within high energy physics, a few dozen papers are published daily. Content is freely accessible online. While other repositories provide search tools and aggregated information, such as author pages or citation counts, most readers access the central arXiv via quick scans of the titles and author names of the daily set of papers. This repository, I argue, is a space of flows (Castells, 1996) populated by things-in-motion that structure the temporalities of scholarly publishing. The space allows for a renegotiation of the nature of the objects that come to constitute a publication. With Foucault (1999), I wonder which kinds of 'author function' HEP physicists construct or lose when publishing their papers on preprint archives. As the 'solid and fundamental unit of the author and the work' constitutes one of the core bases upon which modern knowledge is built and sedimented, HEP makes both the concept of *work* (oeuvre) and the status of the author highly problematic (Foucault, 1999: 205).

Archives are the result of transformations and clashes, and thus are historically, culturally and technologically situated. Building on the metaphor proposed by Traweek, and based on the technologies used in HEP, that is, articles as traces left by subatomic particles, I propose that arXiv has assumed the role of a *detector* used by the community to visualize the passage of such particles and thus make written objects noticed and their authors available for scrutiny. Detectors are huge machines that can identify and make visible subatomic particles coming either from the environment of from an accelerator. In *Archive Fever*, Derrida (1995: 17) maintains that 'archivization produces as much as it records the event'. In this vein, and regardless of the content that is being digitally recorded, the publication of a manuscript on arXiv produces an event – a blip of the detector – that would not be otherwise detected by the

HEP community. Using this metaphor, in this paper I analyze how high energy physicists shoot or refuse to shoot their manuscripts through arXiv as beams of particles produced by an accelerator. They do so in the hope of being recognized as authors, as well as legitimate and productive members of their community. In order to achieve this, high energy physicists need to abide by the way their archive structures the condition of being an author.

The ways in which arXiv organizes time and contributes to negotiations about authorship claims have thus made it central to the community itself. Indeed, as individuals immersed in an open access world based on digital exchange and communication, HEP physicists learn to master the tempos and characteristics of the flows that pass through the archive. It is this circulation of written objects in different regimes of value in space and time (Appadurai, 1988; Daston, 2000) that needs to be followed in order to grasp the archive's cultural and community-structuring significance. Thus the questions of when, where, and with what the detector needs to be hit become urgent. Flows of written objects exist in older scholarly media too: Libraries send out newsletters informing about their latest acquisitions, bookstores show new arrivals on a dedicated table and paper-based preprint archives used billboards to post the latest articles received. Yet two dimensions, one temporal and the other organizational, have modified the role of scholarly object circulation in HEP. First, the digitization of scholarly communication has contributed to an acceleration of time within the community. Some HEP physicists now live in a state of digital immediacy in which the production, circulation and usage of scholarly communication is both accelerated and more central. Of course, media technologies play an important role in the changing speed of community life more generally (Tomlinson, 2007; Virilio, 2006) structuring tempos of work (Munn, 1992). Second, the growth of transnational collaborations that operate particle accelerators and detectors, including hundreds, or even thousands, of physicists, has both slowed down time and created serious problems of authorship. Collaborations list articles' authors in alphabetic order, making individual physicists completely invisible (Galison, 2003). The archive provides high energy physicists with a space for claiming their role within the community through a redefinition of what is a legitimate object of scholarly communication.

This work is based on archival research and semi-structured interviews conducted in 2015 with twenty high energy physicists in Italy and California. The sample includes physicists at different stages of their career: advanced PhD students, postdoctoral scholars, faculty members at various stages, as well as retired physicists. The research was aimed at understanding the history and daily use of a wide set of media, including arXiv and other preprint repositories, as well as the core journal publications in the field. In this paper I focus on material linked to writing, publishing and reading habits in two sub-groups as defined by two categories used by arXiv itself to organize publications: *theorists*, who provide mathematical tools to predict and interpret empirical findings, use the HEP-TH category, and *experimentalists*, who build and operate particle accelerators and detectors, use the HEP-EX category. Both groups use preprint archives as their main communication platforms, but have different writing, publishing, authoring and reading patterns, which in turn lead to different strategies of interaction with the archive. This categorization may oversimplify the nuanced epistemic cultures that compose HEP as well as other forms of physics (Knorr Cetina, 1999; Traweek, 1988). Furthermore, it ignores the different practices of high energy physicists who publish in other, albeit less popular, threads of arXiv, such as phenomenologists or instrumentalists (Larivière et al., 2014; Reyes-Galindo, 2016). Yet a thick analysis of the practices of HEP theorists and experimentalists shows how, within a common digital scholarly communication infrastructure, different communities use preprint archives to meet their different internal communication needs, and how this is tied to evolving issues of authorship and credit attribution.

**Digitizing the archives**

Digital preprint archives have evolved in clear continuity with printed publication formats. Regardless of hegemonic descriptions of digital scholarly communication as 'revolutionary', change in media (and thus in publishing systems) often maintains a balance between continuity and discontinuity (Borgman, 2007). Through cycles of 'remediation', new media technologies evolve out of older media while preserving underlying communication formats and practices; unsurprisingly, old practices seem to fold into new technologies and shape them continuously (Bolter and Grusin, 2000). In the early days of the use of modern computational technologies in science, computerization has been embraced first and foremost in settings in which it could integrate and make more efficient already existing material practices. Rather than representing a rupture, computerization had to do with increased efficiency enacted by transferring existing objects onto a new medium (Agar, 2006). In HEP, the remediation of printed preprint circulation into the digital realm has contributed to the evolution of the archive's role, while preserving some of its sociotechnical characteristics and the 'preprint culture' that sustains it. Understanding its genealogy is key to grasping the transformations fostered by digitization and thus the renewed cultural significance of preprint archives. Contemporary open access platforms for knowledge exchange are often based on pre-existing social norms for exchanges in past models of scientific production. Digitization shapes the spaces of knowledge exchange and circulation, yet builds upon previous cultural characteristics and evolves out of older technologies. For example, a culture of sharing specimens in natural history museums has shaped the foundation of GenBank as an open access database for sharing genetic sequences (Strasser, 2011). Similarly, the emergence of open source synthetic biology mirrors printed model organism newsletters, which were exchanged by biologists in order to share knowledge about the common basic tools of research, such as *arabidopsis* or *drosophila* (Kelty, 2012). Today's 'preprint culture' of high energy physics is the product of a precise cultural and technical past.

Since the end of World War II and the subsequent growth of the field of physics, the practice of exchanging preprints by post, irrespective of the distribution of the same papers via science journals, was institutionalized (Bohlin, 2004; Gentil-Beccot et al., 2010). When they thought that an article was ready to be read by colleagues, physicists would take it to their laboratory's or department's library, for distribution to other libraries. The libraries would keep archives of preprints sent in from around the world. On a bulletin board, often positioned at the entrance of the library, researchers could read the titles and the authors lists of preprints; if researchers were interested in a paper, they could request a copy. This was an expensive practice, and therefore the wealth of a department or laboratory would influence the quantity of preprints it was able to send out, and its reputation influenced the quantity of incoming preprints. The ability to send and receive a high volume of preprints meant that a physicist was positioned within a core area of the discipline, such as an important laboratory or department. For example, the drawers of CERN's archive occupied a full hallway. A physicist's place in one of the core networks that cross the community – a position that is possibly more important than one's record of peer-reviewed publications – was thus crucially expressed by her ability to send and receive preprints. In fact, before the emergence of digital media the role of journals was already peripheral: Physicists did not read journals and rather relied on archives to exchange and retrieve information.

Physics gets old quickly. This epistemic characteristic means that a piece of work can be forgotten as fast as in a few months (Traweek, 1988: 86). Thus there is value in oral communication and preprints, to avoid the long gap between submission to a journal and publication, when an article might already have become less interesting. This has been transposed almost perfectly to the current system, where the digital archive is a faster, centralized, more convenient and reliable way of exchanging papers.

> Often I know in advance what people have been doing. News travel faster than the frontier of knowledge, meaning that physics is now very slow and incremental, with no breakthroughs, so you more or less know or expect what you will see. [TH postdoc][1]

Since the 1970s, preprint exchange has moved to emails and online newsletter first, and eventually to website-based repositories. ArXiv.org, born in 1991 as the aggregation of a number of previously existing online archives and newsletters, represents the core publishing venue for HEP. It is based at Cornell University and operated by a relatively small group of editors financed through grants. Since 1991, arXiv.org has grown to comprise a vast majority (99%) of HEP written output. It has also become the standard for the publication of research results in other sectors of physics, mathematics, computational biology and other disciplines.

As soon as a paper is thought to be ready to be made public, its authors submit it to arXiv. Physicists submit their preprints to the clearly separated and specific sub-section that reflects their disciplinary position, such as HEP-TH for theoretical high energy physics and HEP-EX for experimental high energy physics. The paper is then subjected to a fast review process that may only last for few hours. This includes institutional limitations, as the paper must be sent from a recognized institutional email account and a new author must be endorsed by a recognized user; automated textual analysis screens the content for plagiarism, and that is followed by human moderation (Reyes-Galindo, 2016). Once published, articles are included in a daily digest that aggregates them in a strict chronological order. Only at a later time, in many cases a few weeks after its appearance on arXiv, the same paper might be submitted to a journal to go through peer review, a process that can last several months.

Physics journals seem to have a very high acceptance rate. According to recent estimates, 10,000 papers a year are published by HEP physicists on the archive, and 5,000 formal publications appear in peer-reviewed journals (Larivière et al., 2014; Mele et al., 2006). I will discuss this discrepancy later. Furthermore, HEP papers in a certain sub-field are published in the same few journals. For example, in theory (HEP-TH in the archive) the core journals are *Physical Review Letters*, *Physical Review D*, *Journal of High Energy Physics*, and *Nuclear Physics B*. It is unclear how important it is to publish in a certain journal, at least among the ones that are recognized as 'core' HEP publications. With very few exceptions, these journals struggle to differentiate themselves in terms of prestige, and HEP theorists only acknowledge the superior prestige of *Physical Review Letters*, a journal that publishes short papers of general interest. Thus negotiations around the appropriate journal to send an article to are minimal.

> Some friends in other sciences ask each other, 'in which journal do you publish?' But nobody in particle physics would ask that. [TH postdoc]

When the paper is accepted and published by the journal, often the authors replace the file existing on arXiv with the most updated version, especially if the manuscript has undergone substantial amendments or improvements. In some cases, authors even publish *post-print* versions of a paper, meaning that they update the final, accepted journal paper with a newer and better version: in such cases, the version available on the journal website becomes outdated.

> Diligent people will publish a final version on the archive. [TH faculty]

Both HEP theorists and experimentalists are immersed in a common open access infrastructure based on preprint circulation and on a separation between the spaces where information circulates and assumes internal legitimacy – archives – and those that certify the relevance of knowledge for external actors – peer-reviewed journals. Nevertheless, their epistemic and institutional characteristics structure different forms of interaction with the space of flows created by archives. In different ways, both theorists and experimentalists use arXiv to claim their individual position within their respective community. Theorists live in a state of accelerated time that shapes their reading and publishing practices around precise cycles. Experimentalists resort to tactics that allow them to circumvent the slowed-down time and invisibility they experience as members of large collaborations.

**Theorists: Mastering the tempos of flow**

HEP theorists experience a great deal of individual power over decisions related to scholarly publishing. As part of their education, they learn to use this power to exert control over the timing and spaces of reading and publishing. In the course of their PhDs, young theorists are socialized into the community's communication systems, learning to rely on oral and face-to-face communication for the discussion of knowledge (Traweek, 1988). At the same time, though, they learn to use the archive first as readers, and later as authors. In this process, they need to learn how to deal with and command three different temporal scales: the minute, the day, and the year. In fact, as part of the learning process that introduces physicists to the shape and dynamics of their community, students are trained on the functioning of the archive and gradually learn to look at it on a daily basis, as well as to become part of its flow as authors.

Preprints submitted to and accepted by arXiv are published every weekday at 8 PM EST, listed in order of submission; these number between one and three dozen each day. There are several tools that physicists use to tap into this daily flow. They can open the main website, arXiv.org, or use other technologies that provide the day's list of preprints every morning, such as RSS feed, an email alerting service and a smartphone app called 'arXiv mobile'. Not surprisingly, the latter is especially used by younger physicists, such as advanced PhD students and postdocs, who might use it to browse the daily set of articles while on their morning commutes. These different interfaces allow theorists to browse through the list of the papers published day by day in the HEP-TH subcategory. Students and postdoctoral fellows, who play a fundamental transitional role in physics, must be ready to discuss who published what that day, possibly at the department cafeteria during lunch break. Research on the acceleration of academic life has stressed how junior scholars are particularly exposed to demands for increased productivity that are common in contemporary research. Furthermore, junior scholars tend to be conscious of the need to master specific temporalities (Vostal, 2014; Ylijoki and Mäntylä, 2003). This seems to be particularly evident for postdocs, as shown in the life sciences by Müller (2014). On the other hand, physicists at more advanced stages of their career tend to be less rigorous in checking archives on a daily basis: They seem to become less obsessed with this flow of papers, at least as readers, while recognizing how younger researchers need to live within the archive flux. This is true across the board, and seems to be recognized by theorists and experimentalists alike:

> The kids are passionate, they use the archive on a daily basis. [EX faculty]

Senior physicists also tend to use other archives to browse for useful papers, and perhaps to check citation records if they are evaluating aspiring postdoctoral fellows or researchers in the screening steps before hiring a new colleague. As fast as it has become, for many physicists the digital archive still seems to be slower than other communication practices like word of mouth, in a world in which 'most people know most other people' [TH postdoc]. If a HEP physicist is centrally positioned within the community he or she will be able to rely on oral communication. Indeed, even though they rely on arXiv for the daily flow of preprints, physicists use a different archive for searching for the articles that they need to use in their work. For example, inSPIRE (an evolution of SPIRES, or Stanford Physics Information Retrieval System) is a website that provides an overlay to arXiv: It uses information contained on arXiv, such as the unique numeric identifier that is associated to each and every published article, in order to provide metadata.[2] On inSPIRE, physicists can easily retrieve old articles as well as access data about citations, co-authorship, and articles published in peer-reviewed journals. Yet understanding how physicists interact with the daily flux on the main archive is crucial for grasping the digital immediacy of physics' scholarly communication system. This digital immediacy prevents physicists from analyzing the content of preprints that appear as parts of the flow on the archive. The sheer number of papers published every day makes it impossible to read the content unless one limits attention to specific fields of study, such as supersymmetry or topological string theory. On the other hand, preprints can be useful to claim priority over a piece of knowledge or signal emerging collaborations through multi-authored papers:

> Who the author is is an important piece of information on the archive. [TH faculty]

In the daily flow of articles on arXiv, reading is thus usually limited to the titles and the authors, in a few cases includes the abstracts, and much more rarely extends to an entire article, if downloaded for future reading. Checking the archive on a daily basis teaches physicists what the subjects of the moment are, and being able to discuss the relevant ones proves one's membership in the community.

If the times of browsing the daily flow of preprints are accelerated for readers, successful authorship becomes a matter of timing, too. Theorists need to keep a strict control over two further temporalities fostered by the archive: publishing frequency and instantaneous submission. Publishing frequency means that theoretical physicists need to make their name appear in the archive a suitable number of times per year. This frequency guarantees a visibility that is used to claim a role within the community. The archive not only informs about the content of research projects and about productivity, but also about who is working on what and collaborating with whom, thus making public the kinship relationships that are so crucial within the community (Traweek, 1988). This is especially true for advanced graduate students and postdoctoral scholars, who might identify a specific target frequency, even though this is not related to the perceived quality of their work.

> This seems to be important, not too many, but not too little. Too many means the quality is low, too little you're not committed enough. Maybe the right number is three papers per year? The average might be four or five per year but some people can get away with three or even with two if they're very good. … You need a certain number of papers a year to stay in the game. But how many papers you need to find a truth, that's up to you. [TH postdoc]

While the content of their scholarly output might have been discussed at conferences and chalkboards within their department, HEP theorists shoot their own papers in the beam in the hope that the community will detect those papers an appropriate number of times per year. This accelerated cycle is made possible by epistemic and institutional factors that differentiate the 'cost' of time for theoretical and experimental HEP.

> Experiments can take up to five years to complete, so experimentalists have to choose which problems they work on: Five years is a significant fraction of one's lifetime. But in theory time is cheap. You can work on something for a month, draft a paper and put it on the archive. [TH PhD candidate]

Finally, whereas the majority of the preprints published on arXiv turn into papers published by peer reviewed journals, the importance of this passage is often downplayed. After its publication on the archive, theoretical physicists expect to receive comments from colleagues. In most cases this form of open peer review boils down to a small number of emails, with requests for citations or minor comments. Yet it appears that the peer review process carried out by the journal that receives a manuscript is likely to be even shallower, and indeed the core journals tend to have a very low rejection rate (Bohlin, 2004). Theory postdocs are much more excited and nervous about the appearance of one of their papers on the archive than about its submission to a peer reviewed journal.

> The archive is the most important marketplace …. Once the paper is out there, the main thing is done. The journal publication is the dessert. [TH postdoc]

> The most exciting part of publishing a paper is publishing it on the archive, and also giving talks on it. [TH postdoc]

The last temporality has to do with the functioning of arXiv's submission system. Papers are posted daily and, as stated on arXiv.org, they 'will be entered in the listings in order of receipt on an impartial basis'.[3] Furthermore, physicists know that papers in the top or bottom tier of the day's listing receive more hits and are more likely to be detected. Thus they need to pay attention to the archive's daily deadline: submissions gathered every day before 4 PM EST are made publicly available on that evening and then appear in the RSS or mailing list the next morning. Submitting a paper at 4:01 PM ensures that it will be amongst the first ones in the next group of articles. Physicists race to submit their most important papers at the right moment in order to meet the need for increased visibility, for example when they are submitting a crucial piece of work or an article emerging from an important collaboration. They wait for this internal deadline in the hope that their work will benefit from additional visibility.

> People know there is a correlation: The first or last papers to appear on the day's list get more views …. Should it be random? This has been discussed and someone thinks it's a form of self-selection: One should pay more attention to papers whose authors have put effort into making it to the first part of the list. [TH postdoc]

> When they think it's an important paper, people get ready to push the button at the right moment. [TH postdoc]

HEP theorists thus consciously deal with three temporalities structured by the archive and learn to shoot their beams of papers accordingly. Digital immediacy is at play for the temporalities of reading and discussing, which revolve around a 24-hour scale, and submitting, where the very minute at which one submits a paper to arXiv.org may influence the number of people who detect it. Finally, a year-long scale is what shapes their publishing practices, based on assumptions about the appropriate publishing frequency. By hitting the archive at the right moment and frequency with their beams of papers, theorists claim their role within their peers.

**Experimentalists: Beams of unknown particles**

Experimentalists have little control over the timing of their publications, which is slowed down by the temporal scale of the experiments they run, or over individual authorship claims. They work in large-scale transnational collaborations that design, build, maintain and operate particle accelerators and detectors, among the largest scientific instruments ever built. Each article produced by the collaboration lists as authors all of the hundreds or even thousands of researchers, PhD students, and technicians who contribute to the building and operation of a detector, such as the ALICE experiment at CERN ('A Large Ion Collider Experiment'), that analyzes quark–gluon plasma. Furthermore, in most collaborations names are listed in strict alphabetic order, irrespective of actual contributions. The research, data collection or analysis that leads to an article is carried out by specific groups of physicists, whose names are diluted in the collaboration list. This collectivist practice has been called 'hyperauthorship' (Cronin, 2001). An experimentalist might be listed as an author in hundreds of publications, to only a handful of which she has directly contributed, and only a minority of which she has even read. This can already be evident at initial career stages.

> I have written two articles, published one hundred and thirteen …. I already got to the point where I can no longer list the papers from the collaboration on my CV. [EX postdoc]

While other scholars have discussed the changes that this organizational form brings about for HEP physicists as 'epistemic subjects' (Birnholtz, 2006; Knorr Cetina, 1999), here I am concerned with the role of scholarly communication, and of preprint archives in particular, for individual authors. Articles produced by collaborations do not allow for the attribution of credit related to a physicist's individual contribution to an experiment or measurement, so single authors must find alternative strategies to communicate their direct contribution to a research project, or on which subjects they are working.

Forms of credit attribution based on standard peer-reviewed or preprint articles are excluded for two classes of reasons. First, the temporalities at play in HEP experiments can exceed a single physicist's career, as experiments organized around massive particle accelerators and detectors are conducted on multi-decade temporal frameworks that are beyond the control of any individual physicist. Second, individual physicists affiliated with a collaboration do not have the authority to submit their articles for publication. Collective authorship is mediated by complex political and bureaucratic arrangements over the publishing process. Once physicists have collected data from an experiment they present their results in internal seminars. Once they have drafted a manuscript they must circulate it via internal mailing lists, present it in internal seminars, and eventually submit it to an internal editorial committee. If the committee approves the manuscript, it will also be in charge of deciding where and when to submit, usually on arXiv and contextually to a journal. The final article will list as authors all members of the collaboration in alphabetical order. These processes are governed by internal rules that are based on a lexicon of deliberative democracy adapted to the regulation of matters of publishing and authorship. Often these rules are called 'constitutions', such as in case of the ALICE experiment and its more than 1,300 members (ALICE, 2014). Authorship rights are typically attributed based on a 'labor model' (Biagioli, 2003: Galison, 2003): In most cases, a physicist who joins the collaboration must work on its detector for six months or a year before gaining the right of being listed on each article produced by the collective, and maintains this right for the same amount of time after leaving. Within the lexicon of deliberative democracy adopt by collaborations, these constitutions define the participation in multi-authored articles as 'endorsement'. When an individual researcher decides not to be listed as author on a paper published by the collaboration, it is said the she has withdrawn her endorsement.

According to Foucault, with modernity we came to accept scientific texts as the product of anonymity. The scientific author is diluted within the membership in a collective effort, and her individual name does not stand as the guarantee of truth. According to this view, in modern science the author function fades away, as the credibility of claims about Nature is provided by the scientific method of knowledge production rather than by the person who makes them (Foucault, 1999: 213). STS scholars working on high energy physics have maintained that in this field the author function fades away even in relation to credit attribution (Biagioli, 2003; Traweek, 1988). While traditional ways of constructing an author function are occupied by collectivist arrangements that obscure individual contributions, physicists do not accept the dilution of the author function and struggle to produce and circulate claims related to their individual stance within the field. How, then, is credit attributed besides oral and informal communication?

Individual experimental physicists accumulate symbolic capital through face-to-face communication activities, for example at the chalkboard during internal seminars, where they literally show their peers they can 'get the sums right'. Conference presentations, internal memos, letters of recommendation and gossip are crucial forms of communicating individual credit over a certain measurement or piece of equipment (Galison, 2003).

Beyond such informal methods, HEP experimentalists need to master innovative strategies for establishing their individual contributions to an experiment and role within the community. These physicists use what appear to be liminal, but can actually be quite central, objects in the archive. They shoot out beams of objects that exceed formal articles produced by the collaboration and are not destined for peer review, but that establish their individual identities as authors. These objects take the form of conference proceedings, technical reports, data, white papers or presentations from conferences.

> The only things I upload myself are proceedings. So I have very few articles with some signatures, perhaps 4 to 10. Normally my papers with the collaboration can have up to 1,200. So I have more than 100 articles. Most I haven't written, some I haven't even read …. In the absence of a hierarchy in signatures, people can make themselves more visible by presenting at conferences, or being the first name on a proceeding paper. [EX faculty]

> For me, another form of publication is the white paper that describes the preparation of an experiment. [EX faculty]

This so-called 'grey literature' (Banks, 2006) accounts for part of the missing 5,000 objects that appear on the archive under the banner of HEP and are never published in peer-reviewed journals. It is through these objects that would be considered irrelevant in other disciplines that experimentalists publicly claim credit over their own measurement or over a piece of hardware. For experimental physicists, the archive is the space that allows the detection of these objects within its daily flux.

As members of collectivist organizations, experimental physicists also need to find ways of expressing dissent. In software development, 'forking' is a process in which some developers take a piece of software or a package and start to develop it independently of the original product or community, giving birth to a distinct piece of software. One example is Linux Mint, an operating system forked from Ubuntu, which in turn derives from Debian. The right to fork a product is explicitly allowed by free software licenses such as the GPL, or General Public License (Kelty, 2008). Like members of free software communities, experimentalists work in geographically dispersed collaborations organized through complex bureaucracies, complete with subcommittees and working groups; each team or individual is in charge of modular contributions that are aggregated within a more complex product such as a detector; decisions on when to publish a new dataset or paper is bestowed upon editorial committees that supersede individual contributors. Yet HEP physicists cannot fork a project, as the sheer materiality of a particle detector cannot be replicated by anything smaller than a transnational, multi-billion dollar collaboration.

In the democratic model that underpins their constitutions, HEP collaborations create space for expressing dissent in a number of different forms. Although originally developed in relation to consumer choices, a reference to Hirschman's (1970) theorization of exit, voice and loyalty might be of interest here. Members of a collaboration can have their *voice* heard in internal meeting and boards and thus modify the trajectory of a piece of research with which they disagree. They can show *loyalty* by abiding by the authorship roles and endorse a colleague's manuscript. Or they can temporarily *exit* the collaboration by withdrawing their endorsement and thus becoming non-authors of a particular article that is to be published on arXiv. This form of subtraction is rare, heavily moralized and in some cases explicitly discouraged, as it undermines the force of the argument made in a publication, as well as the consensus within the collaboration (Galison, 2003). Disagreement can also take the form of outspoken dissent, even though it does not mean that dissenting physicists will be expelled from the collaboration. It may also not be noticeable in the archive and often needs to be brought to light by more public forms of action, such as internal discussions or even interviews with the press.

Experimentalists thus need to master different objects and forms of dissent in order to publicly establish their role as individual members of the community. Grey literature becomes central, as it can be detected in the archive as the product of individuals or small groups, because the author function is reinstated. It also allows physicists to overcome the slowed time of arXiv publishing by large-scale collaborations. Finally, non-authorship becomes crucial as a way of expressing dissent over collective choices without breaking the collaboration.

**Conclusions**

High energy physics is still based on an oral culture of face-to-face communication, as well as on kinship relationships that are only partially captured within its formal communication system. Yet preprint archives have assumed a central role within the field as the spaces where belonging and credit are constructed and confirmed. The role of such archives has evolved, as a result of the growth of HEP to include thousands of researchers organized in transnational collaborations and networks and joined by a global culture, as well as through processes of remediation that have integrated past practices into digital media. In HEP, preprint archives are thus central to processes of boundary construction and maintainance, as well as to credit and authorship negotiations. In a culture where preprint archives are pervasive and dominant, practices to affirm one's individuality can vary greatly, but often have to do with a continuous flow of papers that pass through the main preprint archive used by HEP physicists, arXiv. Thus, navigating this space of flows has emerged as a central feature of contemporary high energy physics. In this regard, arXiv can be seen as an ephemeral space whose purpose is not solely the preservation of a record of scholarly publications or the certification of knowledge. Following Derrida, I propose that arXiv, not unlike other archives, produces as well as records events. The archive is a detector that HEP physicists need to hit with their beams of papers, at the right time and with the right objects, in order to signal their existence and claim credit for what they have done.

I have described how two subsets of HEP physicists, characterized by different epistemic cultures and institutional arrangements, publish their papers on arXiv. Theorists experience full control over their publishing practices. They learn to master the tempos shaped by arXiv in order to be recognized as productive and legitimate members of their community. On the other hand, as members of collectivist organizations, experimentalists have little control over publishing decisions and experience diluted authorship. They exploit arXiv to render visible scholarly objects such as grey literature, which allow them to claim credit as individual authors.

This leads to a number of further questions. First, as many open access advocates think that high energy physics is a good model for the scholarly communication system of the future, what would it mean for other disciplines or scholarly cultures to embrace preprint archives as their core communication infrastructure? The accelerated pace of production is a common feature of contemporary scholarship. But within this shared framework, many epistemic and institutional factors play a role. How are the tempos and authorship practices of the social sciences or biology shaped by the adoption, as hegemonic venues for scholarly communication, of services such as academia.edu, SSRN or bioRxiv, with their specific constraints and affordances? As digital platforms for preprint publication are adopted by scores of young researchers in many scholarly fields, these services are increasingly becoming the spaces in which credit and authorship are claimed, and the boundaries of scholarly communities negotiated. Studying their technical affordances and political structures may be crucial for understanding the evolution of contemporary scholarship. Second, the growth of massive collaborative research in other disciplines, for example in 'omic' sciences such as genomics or proteomics, will push other scientists to figure out ways of claiming individual authorship over snippets of their collaboration's output – for example, by publishing objects now considered grey literature, such as datasets or white papers. Will digital preprint platforms legitimate the emergence and circulation of new scholarly objects through which credit can be claimed? As in high energy physics, these objects may evolve as either complementary or alternative to the core object of modern scholarly communication, the peer reviewed journal article. Finally, how is scientific dissent expressed through non-authorship? Democratic theory and free software studies suggest that forms of collective authorship may sanction and at the same time foster forms of withdrawal and non-participation, and an exploration of the significance of these practices might be overdue (Casemajor et al., 2015). Like free software, high energy physics organizes dissent through specific governance models. The role that preprint archives will assume in these fields will depend on the different disciplinary cultures, local and epistemic, as well as the different social and institutional arrangements at play. Studying high energy physics thus might change the way we think about the role of preprint archives in scholarship, as well as transform the meaning of concepts such as publication, author, or access.

**Acknowledgments**

I would like to thank Mario Biagioli, Jonathan Eisen, Alexandra Lippman and Sergio Sismondo for their help and support.

**Notes**

1. Interviewees are identified as theorists (TH) or experimentalists (EX) as well as by their academic position (PhD candidate, postdoc, or faculty member).

2. http://inspirehep.net (accessed 21 June 2016).

3. http://arxiv.org/help/general (accessed 21 June 2016).

**Author biography**

Alessandro Delfanti is assistant professor of Culture and New Media at the University of Toronto. He is the author of *Biohackers: The Politics of Open Science* (Pluto Press 2013).

**Please cite as follows**

Delfanti A (2016) Beams of particles and papers. How digital preprint archives shape authorship and credit. *Social Studies of Science*.